\documentstyle[aps,multicol,epsfig,prl,floats]{revtex}
\begin{document}
\draft
\twocolumn[\hsize\textwidth\columnwidth\hsize\csname @twocolumnfalse\endcsname
\author{Yu. B. Khavin and M. E. Gershenson}
\address{Rutgers University,  Serin Physics Laboratory, \\
Piscataway, NJ 08854-8019}
\author{A.L. Bogdanov}
\address{Lund University, Department of Solid State Physics, \\
Nanometer Science Consortium, S-221 00 Lund, Sweden}
\date{\today }
\title{Decoherence and the Thouless Crossover in One-Dimensional Conductors}
\maketitle

\begin{abstract}
The temperature and magnetic-field dependences of the resistance of
one-dimensional (1D) conductors have been studied in the vicinity of the
Thouless crossover. We find that on the weak localization (WL) side of the
crossover, these dependences are consistent with the theory of quantum
corrections to the resistance, and the phase breaking is due to the
quasi-elastic electron-electron interactions (the Nyquist
noise). The temperature dependence of the phase coherence time $\tau
_\varphi $ does not saturate, and the quasiparticle states remain well
defined over the whole WL temperature range. This fact, as well as
observation of the Thouless crossover in 1D samples, argues against the idea
of intrinsic decoherence by zero-point fluctuations of the electrons
(Mohanty {\it et al}., Phys.Rev.Lett. {\bf 78}, 3366 (1997)). We believe that frequently
observed saturation of $\tau _\varphi (T)$ is caused by the external
microwave noise.
\end{abstract}


]

Recently intrinsic decoherence in disordered conductors has been proposed%
\cite{webb1}. It has been noted that many distinct experiments indicate a
saturation of the phase coherence length $L_\varphi $ at low temperatures.
The authors of Ref.\cite{webb1} suggested that this saturation is due to
zero-point fluctuations of the electrons.

In this Letter, we present new data on the temperature dependence of $L_\varphi $ in
one-dimensional (1D) conductors in the vicinity of the Thouless crossover.
The experimental values of $L_\varphi $ are very well described by the
theory of decoherence due to the quasi-elastic electron-electron scattering%
\cite{alt1}. The temperature dependence of $L_\varphi $ {\it does not
saturate }down to the crossover temperature, and the maximum experimental
values of the phase coherence time $\tau _\varphi $ exceed the limit
suggested in \cite{webb1} by a factor of 50. This fact, as well as
previously reported observation of the temperature-driven Thouless 
crossover in these samples%
\cite{gersh1,gersh2,kha}, {\it contradicts} the idea of decoherence by the
zero-point fluctuations of the electromagnetic environment. We attribute the
frequently observed saturation of $L_\varphi $ to the phase breaking due to
the {\it external} high-frequency noise\cite{alt2}. Our samples are much
less ''sensitive'' to the noise-induced decoherence because of a very
high resistance, by a factor of $\sim 10^4$ greater than that for the
Au wires studied in \cite{webb1}.

The temperature and magnetic field dependences of the resistance have been
measured for sub-micron-wide ''wires'' fabricated from the $\delta $-doped
GaAs structures. Similar structures have been used to study the crossover
from weak localization (WL) to strong localization (SL) in 1D conductors 
\cite{gersh1,gersh2,kha}. A single $\delta $-doped layer with concentration
of Si donors $N_D=5\times 10^{12}cm^{-2}$ is $0.1$ $\mu m$ beneath the
surface of an undoped GaAs. The 1D wires were fabricated by electron beam
lithography and deep ion etching. The samples cosist of up to 500 wires
connected in parallel; the length $L$ of each wire is 500 $\mu m.$ A
50-$nm$-thin silver film deposited on top of the structure was used as a
''gate'' electrode: the electron concentration $n$ and the resistance of the
samples can be ''tuned'' by varying the gate voltage $V_g$ (for more
details, see \cite{kha}).

We discuss the results for a typical sample consisting of 360 wires of
effective width $W$ = 0.05 $\mu m$. The values of $W$, obtained
from the sample resistance, are in accord with the estimate of $W$ from the
analysis of the WL magnetoresistance; it has been also verified that $W$
does not depend on $V_g$ \cite{kha}. Parameters of the sample at three
different values of $V_g$ are listed in Table 1. The mean
free path of electrons $l$ increases with $n$ from $17$ $nm$ to $58$ $nm$ ($%
k_Fl\approx 6\div 30$, where $k_F$ is the Fermi wave number). A relatively
high concentration of carriers ensures that the number of occupied 1D
sub-bands $N$ is large ($\sim 10$), and the localization length $\xi \sim Nl$ is much
greater than $l$. Two-dimensional expressions for the diffusion constant $D$
and the density of states $\nu $ are used because the broadening of energy
levels due to the elastic scattering is greater than the spacing between the
1D sub-bands.

\smallskip
Table 1
\smallskip

\begin{tabular}{|c|c|c|c|c|}
\hline
$V_g,V$ & $
\begin{array}{c}
n \\ 
\times 10^{-12}cm^{-2}
\end{array}
$ & $
\begin{array}{c}
D(10K) \\ 
cm^2/s
\end{array}
$ & $\xi ,\mu m$ & $
\begin{array}{c}
T_0(H=0) \\ 
K
\end{array}
$ \\ \hline
+0.7 & 4 & 250 & 1.1 & 0.7 \\ \hline
0 & 2.7 & 90 & 0.35 & 2.2 \\ \hline
-0.35 & 2 & 52 & 0.15 & 4.6 \\ \hline
\end{tabular}

\bigskip

The temperature dependences of the resistance for different $V_g$ are shown
in Fig. 1. As it will be shown below, behavior of $R$($T$) at high temperatures 
is consistent with the
theory of quantum corrections to the resistance (for a review, see \cite
{alt3}). The crossover from weak to strong localization emerges with
decreasing the temperature. The activation-type temperature dependence of
the resistance has been observed on the SL side of the crossover%
\cite{gersh1,gersh2,kha}. The crossover temperature $T_0$ that corresponds
to the hopping activation energy in the SL regime is shown in Fig. 1 with
arrows.

\begin{figure}[ht]
\epsfig{file=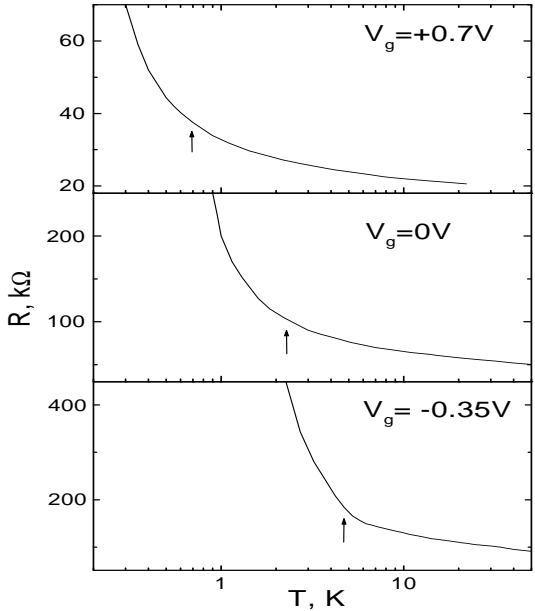, height=0.5\textwidth, width=0.5\textwidth}
\caption{Temperature dependences of the resistance at different $V_g$. The
crossover temperatures $T_0$ are shown with arrows.}
\label{Fig.1}
\end{figure}

The phase coherence length $L_\varphi $ has been estimated from the WL 
magnetoresistance. Experiments with 1D conductors \cite
{tho1,wi1,echt1} have demonstrated that phase breaking in 1D
conductors at low temperatures is governed by the quasi-elastic electron-electron
collisions, which is equivalent to decoherence by the equilibrium 
Nyquist-Johnson noise \cite
{alt1}. The procedure of extraction of $L_\varphi $ from the WL
magnetoresistance in this case has been described in detail in \cite{echt1}.
The high-temperature ($T$ $>T_0$) magnetoresistance has been fitted with the
magnetic field dependence of the 1D WL correction to the resistance \cite
{alt1,alt3}: 
\begin{equation}
\begin{array}{c}
\frac{R(H)-R(0)}{R(0)}=\sqrt{2}\frac{e^2}{\pi \hbar \sigma _1}L_\varphi
f\left[ 2\left( \frac{L_\varphi }{L_{\varphi 0}(H)}\right) ^2\right] , \\ 
f(x)=-\frac{Ai(x)}{Ai^{^{\prime }}(x)}.
\end{array}
\end{equation}
Here $\sigma _1$ is the inverse resistance of a wire per unit length ($%
\sigma _1$ $=L/R(0)N$ for $N$ wires of the length $L$ connected in
parallel), $Ai(x)$ is the Airy function. The phase coherence length $%
L_\varphi =\sqrt{D\tau _\varphi}$ in Eq.1 is determined by the Nyquist time $%
\tau _\varphi$ \cite{alt1}. The length $L_{\varphi 0}(H)=\sqrt{D\tau
_{\varphi 0}(H)}$ combines phase breaking due to the magnetic field ($\tau
_H $) and the {\it strongly-inelastic} scattering ($\tau _{\varphi 0}$): $%
\tau _{\varphi 0}(H)^{-1}=\tau _{\varphi 0}^{-1}+\tau _H^{-1}$. It is noteworthy
that Eq.1 differs from the expression for the 1D WL magnetoresistance \cite
{alt4} which is valid only for the {\it strongly-inelastic} phase breaking.
The latter expression for the 1D WL magnetoresistance, which is often used 
for fitting the experimental data (see,
e.g. \cite{webb1,tho1,wi1}), does not hold if the {\it quasi-elastic} Nyquist
phase breaking is the dominant decoherence mechanism.

For a ''diffusive'' ($W>>l$) 1D wire in a perpendicular magnetic field\cite
{alt4}, 
\begin{equation}
\tau _H=\frac{12L_H^4}{DW^2},
\end{equation}
where $L_H=\sqrt{\hbar c/2eH}$ is the magnetic length. However, if $W\leq l,$
$\tau _H$ is larger than the estimate (2) due to the flux cancellation effect%
\cite{dud,ben}. Numerical calculations \cite{ben} show that $\tau _H$
exceeds the estimate (2) by factor $2.5$ at $l=W$ (this case corresponds to $%
V_g$ = 0.7 V), by 1.2 at $l=0.5W$ ($V_g$ = 0 V), and the difference is
negligible for $l=0.3W$ ($V_g$ = -0.35 V). 
\begin{figure}[ht]
\epsfig{file=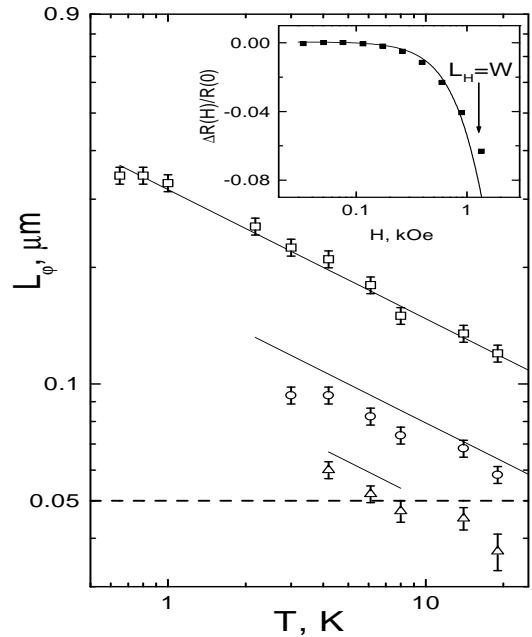, height=0.5\textwidth, width=0.5\textwidth}
\caption{The phase coherence length versus temperature at different $V_g$: $%
\Box $ - $+0.7V$; $\bigcirc $ - $0V$; $\bigtriangleup $ - $-0.35V.$ Solid
lines - Eq.3 calculated for a fixed $D=D\left( T\simeq 4T_0\right) $ ; the
dashed line - Eq.6. The insert shows the magnetoresistance at $T=8K,$ $%
V_g=0V $; the solid line - Eq.1.}
\label{Fig.2}
\end{figure}
The observed magnetoresistance is well described by Eq.1 over the whole
range of fields where the Eq.1 is applicable, $L_H>W$ (the insert in Fig.2).
For the samples studied, strongly-inelastic decoherence processes can be
neglected at $T$ 
\mbox{$<$}
30 $K$ [$L_\varphi <<L_{\varphi 0}(H=0)$], and the {\it only} fitting
parameter is the Nyquist length $L_\varphi $. The temperature dependences of 
$L_\varphi $ (Fig. 2) are in a good agreement with the theoretical result%
\cite{alt1,alt3}: 
\begin{equation}
L_\varphi (T)=\left( \frac{\hbar^2 D\sigma _1}{\sqrt{2}e^2k_BT}\right)
^{1/3}\propto T^{-1/3}
\end{equation}
over the whole temperature range that corresponds to the WL regime. The
dependences (3) (solid lines in Fig.2) are extended down to the crossover
temperature.

As $T$ approaches $T_0$, the dependence $L_\varphi (T)$ is flattened out.
For all the samples studied, $L_\varphi $ at the crossover temperature is
2-3 times smaller than the 1D localization length $\xi =\pi \hbar \sigma _1/{%
e^2}$. The scaling theory of localization\cite{th} predicts that the
crossover occurs when $L_\varphi $ becomes comparable with $\xi $. However,
one should not expect to observe the exact equality $L_\varphi $ = $\xi $ at 
$T_0$, because the crossover is due to {\it both} localization and
interaction effects. The resistance of a wire segment of a length of $\xi $ 
for this sample is $\sim20$ $k\Omega$ at $T=T_0$, which is also consistent with the quantum
resistance $h/e^2$ expected at the crossover\cite{th}.

The temperature dependences of the resistance in the WL regime are also well
described by the theory of quantum corrections. Figure 3 shows $R(T)$ for $V_g=+0.7V$ 
in two cases: {\it a)} at $H=0$, when both localization and
interaction effects contribute to $R(T)$, and {\it b)} at $H=17$ kOe, when the
temperature dependence of the WL correction is completely suppressed ($L_H<l$%
). The theoretical expression for $R(T)$ due to the {\it first-order}
quantum corrections can be written for $H=0$ as follows\cite{alt3}: 
\begin{equation}
\frac{R(T)-R_0}{R_0}=\frac{e^2}{\pi ^2\hbar \sigma _1}\left[ \sqrt{2}\pi \cdot
1.37L_\varphi (T)+4.91\alpha L_T(T)\right].
\end{equation}
The first term in the square brackets is the asymptotic form of the WL
correction (for our samples $L_\varphi <<L_{\varphi 0}(H=0)$ at $T<$ 30 K,
and $f(x)\approx 1.37$ in Eq.1), the second term is the interaction
correction, $L_T=\sqrt{\hbar D/k_BT}$ is the thermal length. Two
parameters have been used for fitting the experimental data at $H=0$: $R_0$
= 16.5 k$\Omega $ and the screening factor $\alpha=0.37$. In strong fields,
only the interaction term in Eq.4 with the same value of $\alpha$ has been
used for fitting $R(T)$\cite{alt3}: we do not expect modification of $\alpha 
$ in this magnetic field because the Zeeman splitting $g\mu _BH\ll k_BT$
at $T>1K$ ($g$-factor is $\sim 0.4$ for the conduction band in GaAs). The best agreement
with experiment at $H=17$ $kOe$ was obtained with a larger $R_0$ $=17.5$ $%
k\Omega $. We believe that the increase of $R_0$ is due to the classical
positive magnetoresistance: $\left( \omega _c\tau \right) ^2$ $=0.1 $ at $%
H=17$ $kOe$ ($\omega _c$ is the cyclotron frequency, $\tau $ is the elastic
scattering time). 
\begin{figure}[ht]
\epsfig{file=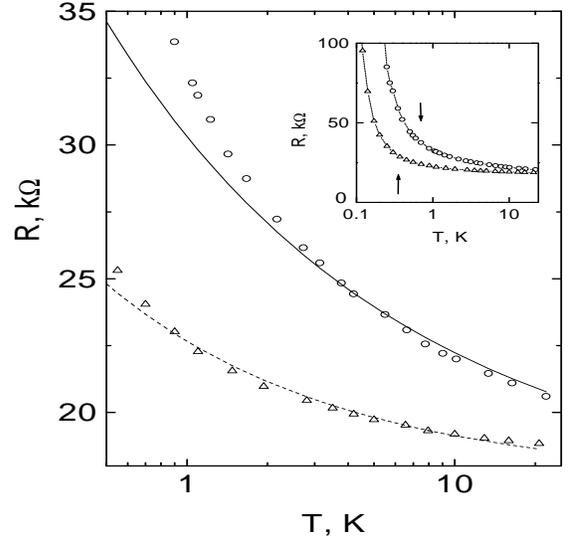, height=0.45\textwidth, width=0.45\textwidth}
\caption{$R(T)$ at $V_g=+0.7V$ for $H=0$ ($\bigcirc )$ and $H=17kOe$ $%
(\bigtriangleup )$. The solid and dashed curves are the theoretical fits
(see the text). The insert shows the same dependences over a larger $T$
range; the crossover temperature $T_0$ is shown by arrows. The solid curves
in the insert are guides to the eye.}
\label{Fig.3}
\end{figure}

An excellent agreement with the theory of quantum corrections is observed
down to $T\approx 3T_0$; at lower temperatures, the higher-order terms of
both localization and interaction contributions to $R(T)$ must be taken into
account. As it has been shown earlier\cite{gersh1}, the shift of the
Thouless crossover toward lower temperatures in strong magnetic fields $%
(L_H\ll \sqrt{\xi W})$ is accompanied with {\it doubling} of the
localization length\cite{ef}, and {\it halving} of the hopping activation
energy in the SL regime.

The observed dependences $L_\varphi (T)$ {\it oppose} the idea of the
decoherence due to zero-point fluctuations of the electrons. 
Indeed, the following expression for the ''cut-off''
decoherence time $\tau _0$ has been obtained for a 1D wire fabricated from a
two-dimensional electron gas (2DEG)\cite{webb1}: 
\begin{equation}
\tau _0=\left( \frac{\pi \hbar ^2}{m^{*}}\cdot \frac{\sigma _1}{e^2D^{3/2}}%
\right) ^2,
\end{equation}
where $m^{*}$ is the effective electron mass. The corresponding length scale 
$L_0$ {\it coincides} with the width of the wire: 
\begin{equation}
L_0=\left( D\tau _0\right) ^{1/2}=\frac{\sigma _1W}{e^2\nu _{1D}D}=W.
\end{equation}
Here $\nu _{1D}=m^{*}W/\pi \hbar ^2$ is the 1D density of electron
states. The existance of the "cut-off" phase coherence length $L_0=W$ 
would imply that {\it a)}
narrow channels fabricated from 2DEG {\it cannot} demonstrate 1D quantum
corrections to the conductivity, and {\it b)} the localization-induced crossover 
{\it should not} be observed in such channels (in this case $L_\varphi $ is always 
much smaller than $\xi \gg W$). This would also preclude observation of the {\it %
interaction-driven} 1D crossover with decreasing the temperature. Indeed, as
soon as $\tau _\varphi $ approaches $\tau _0$, the broadening of the
electron energy levels, $\hbar /\tau _\varphi$, becomes
temperature-independent. The Fermi-liquid description of quasiparticle states 
should also break down at $\hbar /\tau _0>k_BT$ .

Both consequences of Eq.6 contradict our data: we observe the Thouless
crossover in 1D conductors with the ratio $\xi /W$ as large as 16\cite{kha},
and the temperature dependence of $L_\varphi $ does not saturate down to $%
T_0 $. The phase coherence length near the crossover exceeds the estimate
(6) by a factor of ~$\sim $ 7 for $V_g=0.7$ $V $; hence $\tau _\varphi $
exceeds $\tau _0$ by a factor of 50. Similar discrepancy between $\tau
_\varphi $ and estimate (6) has been observed recently in Ref. \cite{link1}.

What is the reason for frequently observed saturation of $L_\varphi (T)$? 
We believe that this saturation is
due to phase breaking by the external microwave electromagnetic
noise. The phase coherence is destroyed most efficiently by spectral
components of this noise with the frequency $\omega \sim \tau _\varphi ^{-1}$%
\cite{alt2,vit1}. The field amplitude $E_\varphi $, which is sufficient
for phase breaking at the time scale $\sim \tau _\varphi $, can be estimated
as $E_\varphi \simeq \hbar /eL_\varphi \tau _\varphi \propto \tau _\varphi
^{-3/2}$ \cite{alt2}. The electric field $E_\varphi $ is a linear function
of $T$ for 1D conductors with the dominant Nyquist phase breaking. The
following noise power is required for saturation of the $\tau _\varphi
(T) $ dependence below some temperature $T$: 
\begin{equation}
P_\varphi =%
{(L\cdot E_\varphi )^2 \overwithdelims.. 2R}
=R%
{ek_BT \overwithdelims() \hbar }
^2.
\end{equation}
This power is proportional to the total resistance $R$ of a wire. The
Nyquist time is very short ($<5\cdot 10^{-12}$ s) in our highly-resistive
samples ($R(4K)\simeq 9M\Omega $ for a single wire at $V_g=0.7V$ ). As a
result, the noise power required for phase breaking at this time scale is
rather large: $P_\varphi \sim 10^{-9}\div 10^{-8}W$. However, for 1D Au
wires with much smaller $%
R\simeq 0.3\div 1.8k\Omega $ \cite{webb1}, $P_\varphi $ is only $1\cdot
10^{-15}\div 1\cdot 10^{-14}W$ at $T=0.1K$. This $P_\varphi $ is%
{\it \ smaller} than the power dissipated by the dc measuring current ( $4\cdot
10^{-14}\div 7\cdot 10^{-13}W)$, which has been experimantally proven not to heat 
the Au wires down to $T=40mK$ \cite{webb2}. As
it is shown in Ref.\cite{aag}, balancing of the incoming microwave power $P_\varphi $ by
the outcoming power due to the hot-electron outdiffusion into
the ''cold'' leads should result in a {\it negligible} rise of the electron
temperature for any 1D conductor with $R\ll h/e^2=25.8k\Omega $. In this
situation, the microwave noise can efficiently {\it destroy} the phase coherence of
the electron wavefunction {\it without} heating the electron gas. This
explains why a well-pronounced temperature dependence of the resistance (due
to the interaction effects) has been observed at temperatures where the
dependence $L_\varphi (T)$ was already completely saturated\cite{webb1}.

To summarize, we show that both the temperature and magnetic field
dependences of the resistance of 1D conductors are well described by the
theory of weak localization and interaction effects on the ''metallic'' side
of the Thouless crossover. The observed temperature dependence of the phase
coherence length is consistent with the Nyquist decoherence mechanism\cite
{alt1} and does not saturate down to the crossover temperature $T_0$. Our
data indicate that the quasiparticle description holds over the {\it whole}
WL temperature range. In
the vicinity of $T_0$, the experimental values of $\tau _\varphi $ exceed $%
\tau _0$ by a factor of 50. This fact, as well as observation of the
Thouless crossover in our 1D samples, {\it argues against }the idea of
decoherence by zero-point fluctuations of the electrons%
\cite{webb1}. Frequently observed low-temperature saturation of the $%
L_\varphi (T)$ dependence can be attributed to decoherence due to the
external electromagnetic noise.

It is our pleasure to thank B. L. Altshuler and I. L. Aleiner for illuminating
discussions. We are grateful to B. K. Medvedev for fabrication of the $%
\delta $-doped layers.

\end{document}